\documentclass[aps,pra,preprint,showpacs,groupedaddress]{revtex4}
\newcommand{\be}{\begin{equation}}
\newcommand{\ee}{\end{equation}}

\begin{document}

\preprint{}

\title{Unitary relation for the time-dependent $SU(1,1)$ systems}

\author{Dae-Yup Song}
\affiliation{Department of Physics, University of Florida,
Gainesville, FL 32611, USA\\ and Department of Physics, Sunchon
National University, Suncheon 540-742, Korea\footnote[2]{Permanent
address}}

\date{\today}

\begin{abstract}
The system whose Hamiltonian is a linear combination of the
generators of $SU(1,1)$ group with time-dependent coefficients is
studied. It is shown that there is a unitary relation between the
system and a system whose Hamiltonian is simply proportional to
the generator of the compact subgroup of the $SU(1,1)$. The
unitary relation is described by the classical solutions of a
time-dependent (harmonic) oscillator. Making use of the relation,
the wave functions satisfying the Schr\"{o}dinger equation are
given for a general unitary representation in terms of the matrix
elements of a finite group transformation (Bargmann function). The
wave functions of the harmonic oscillator with an inverse-square
potential is studied in detail, and it is shown that, through an
integral, the model provides a way of deriving the Bargmann
function for the representation of positive discrete series of the
$SU(1,1)$.

\end{abstract}


\maketitle

\section{Introduction}

Group theoretical methods could be useful in analyzing physical
systems, and particularly the $su(1,1)$-type algebraic structure
is known to appear in many quantum systems
\cite{Wybourne,IKG,Eberly}. The time-dependent quadratic system (a
generalized harmonic oscillator) \cite{Song99} is a realization of
particular representations of the $SU(1,1)$ group. The evolution
operator and transition probabilities of the harmonic oscillator
with a time-dependent frequency have been known in terms of
classical solutions of the oscillator \cite{Seleznyova}. The wave
functions of the quadratic systems \cite{Song99,Seleznyova}, if
the centers of probability distributions of the functions remain
at the origin of the space coordinate, are closely related to the
$SU(1,1)$ coherent states of Perelomov \cite{Perelomov} which are
obtained by applying displacement-type elements of the group on a
fiducial vector in a representation space.

Unitary transformation methods have long been recognized as a
useful tool in finding the wave functions of the coherent systems
\cite{Stoler} and of the generalized harmonic oscillators
\cite{Seleznyova,Unihar,SongUni}. Through a unitary transformation
method, the complete set of wave functions for a general quadratic
system has been given in terms of the classical solutions of the
system \cite{SongUni,Song99}, and the fact that wave functions are
described by the classical solutions can be clearly understood
from the path integral approach for this system \cite{Song99}. On
the other hand, it turns out that the unitary transformation for a
time-dependent quadratic system can be used for the same quadratic
system with an inverse-square potential to give the wave functions
\cite{SongUni}. Indeed, $su(1,1)$ symmetry has been noticed in the
model of the inverse-square potential \cite{inverse}, and the
symmetry has been used to find the stationary wave functions for
the case of a constant Hamiltonian \cite{Perelbook}, which would
imply that a unitary transformation method may be applicable for
general time-dependent $SU(1,1)$ systems.

In this paper, we will consider the system which is described by
the Hamiltonian
 \be
 H=\hbar[ A_0(t)K_0 +A_1(t)K_1+4a(t) K_2) ]+\beta(t),
 \ee
where $K_0,K_1,K_2$ satisfying the commutation relations
 \be
[K_1,K_2]=-iK_0,~~~[K_2,K_0]=iK_1,~~~[K_0,K_1]=iK_2,
 \ee
are the Hermitian generators of the $SU(1,1)$ group and
$A_0(t),A_1(t),a(t),\beta(t)$ are real functions of time, $t$,
with $A_0(t)\neq A_1(t)$. This system has long been considered
\cite{Gerry,DDT,IKG}, and it has been suggested that solutions of
a classical equation of motion might be used in describing the
wave functions \cite{Gerry89}. Since $\beta(t)$ can be understood
as a result of a simple unitary transformation which does not
depend on the generators (see, e.g., Ref.~\cite{Song99}), from now
on we will take $\beta(t)=0$. As an extension of the unitary
relation in the quadratic systems \cite{Seleznyova,SongUni}, we
will give the unitary transformation which relates the system of
$H$ and the system described by
 \be
 H_0=2\hbar w_c K_0,
 \ee
where $w_c$ is a positive constant. The unitary transformation is
described by the classical solutions of a time-dependent
(harmonic) oscillator. With a choice of the realizations of the
generators in terms of the canonical coordinates, the relation we
will give becomes the known one of the quadratic systems
\cite{SongUni}; Due to the non-uniqueness in realizing the
generators, however, the relation in $SU(1,1)$ is more general
than that in the quadratic system even for the representations of
the $SU(1,1)$ which correspond to the quadratic system.

In the next section, we will give the unitary relation between the
systems of $H$ and of $H_0$, and the relation will be discussed in
some explicit realizations. In Sec.~III, making use of the unitary
relation, the wave functions satisfying the Schr\"{o}dinger
equation will be given in terms of the matrix elements of a finite
group transformation (Bargmann function) which in turn will be
determined by the classical solutions of a (harmonic) oscillator.
In Sec.~IV, the representations which correspond to harmonic
oscillator systems will be studied, and other expressions of the
Bargmann function for these representations will be given, which
generalizes the known results on transition probabilities
\cite{Seleznyova}. It will be further shown that the wave
functions of the system of $H_0$ obtained through the unitary
transformation can be written in a simple form. In Sec.~V, the
wave functions of the quadratic system with an inverse-square
interaction will be studied, while the set of the wave functions
gives a representation space of the positive discrete series $D^+
(k)$ which is one of the unitary irreducible representations
(UIRs) of the $SU(1,1)$ group. It will be shown that the wave
functions of the quadratic system with an inverse-square potential
could be used to find the Bargmann function of $D^+ (k)$ through
an integral. The last section will be devoted to the discussions,
and an appendix is added to reveal the equivalent expressions of
the Bargmann function.

\section{A unitary relation}
For the description of the unitary relation, we introduce $M(t)$
as
 \be
 M(t)={2w_c \over A_0(t)-A_1(t)},
 \ee
and $u(t),v(t)$ as the two real, linearly independent solutions of
the second-order differential equation:
 \be
 \ddot{y}+{\dot{M}(t) \over M(t) }\dot{y}
 +[{1 \over 4}(A_0^2-A_1^2)-4a^2+{2\over M}{d\over dt} (Ma)]y=0.
 \ee
For ${1 \over 4}(A_0^2-A_1^2)-4a^2+{2\over M}{d\over dt} (Ma)> 0$,
this is an equation of motion of a generalized harmonic oscillator
\cite{Song99}. By defining $\rho(t)$ and a time-constant $\Omega$,
which are positive, as
 \be
 \rho(t)=\sqrt{u^2(t)+v^2(t)}
 \ee
 \be
 \Omega=M(t)[u(t)\dot{v}(t)-\dot{u}(t)v(t)],
 \ee
and a real function of $t$, $\tau(t)$, through the relation
 \be
 e^{i\tau}={u+iv \over \rho},
\ee
 one may find that the unitary operator
 \begin{eqnarray}
 U&=&\exp\left[i{M\over 2w_c}\left({\dot{\rho}\over \rho}+2a\right)
    \left( e^{2iw_c t}K_+ +e^{-2iw_c t}K_- + 2K_0\right)\right]
    \cr
 & &\times \exp\left[\ln\left(\sqrt{w_c \over \Omega}\rho\right)\left(
    e^{2iw_c t}K_+ -e^{-2iw_c t}K_- \right)\right]
    \exp\left[2i(w_c t -\tau)K_0\right]
 \end{eqnarray}
satisfies the relation
 \be
 U\left(-i\hbar{\partial \over \partial t}+H_0\right)U^\dagger=
  -i\hbar{\partial \over \partial t}+H.
 \ee
In Eq.~(9), $K_+,K_-$ are defined as
 \be
 K_+=e^{-2iw_c t}\left( K_1+iK_2 \right),~~~~~K_-=K_+^\dagger,
 \ee
so that
 \be
{d\over dt}\left( e^{2iw_c t}K_+\right)=0, ~~~~~{d\over
dt}\left(e^{-2iw_c t}K_-\right)=0.
 \ee
The generators $K_0,~K_+,~K_-$ then satisfy the commutation
relations
 \be
 [K_0, K_{\pm}]=\pm K_{\pm},~~~[K_+,K_-]=-2K_0.
 \ee
By making use of the commutation relations in Eq.~(13), with the
fact that
 \be
 {d\over dt}(M\dot{\rho})-{\Omega^2 \over M\rho^3}
 +M\left[{1 \over 4}(A_0^2-A_1^2)-4a^2+{2\over M}{d\over dt}
(Ma)\right]\rho=0,
 \ee
one can explicitly verify the relation of Eq.~(10) \cite{comment}.

The Casimir operation $C$,
 \be
C=K_0^2 -K_1^2-K_2^2=K_0^2- {1\over 2}\left( K_+ K_- +K_- K_+
\right),
 \ee
is used in characterizing  the UIRs of the $SU(1,1)$ group which
are all infinite-dimensional. If we re-parameterize the
eigenvalues of $C$ as $k(k-1)$,  it has been known that, for both
cases of $k=1/4$ and $k=3/4$, the $su(1,1)$ algebra can be
realized by the operators of a quadratic system; If $L_0,L_1,L_2$
is written as
 \be
 L_0 = {1\over 4\hbar}\left( {p^2\over w_c} +w_c x^2\right),~~~
 L_1 = {1\over 4\hbar}\left( -{p^2\over w_c} +w_c x^2\right),~~~
 L_2 = -{1\over 4\hbar}(xp+px),
 \ee
with the commutation relation $[x,p]=i\hbar$, one can find that
$\{L_0,L_1,L_2\}$ can be a basis of the $su(1,1)$ algebra with
$C=-(3/16)I$. If this expression of the generators of the
$SU(1,1)$ group is plugged into Eqs.~(9,10), one can find the
relation
 \begin{eqnarray}
 U_L\left( -i\hbar{\partial \over \partial t}
 + {1\over 2}(p^2+w_0^2x^2)\right)U_L^\dagger &=&
 -i\hbar{\partial \over \partial t} +{p^2 \over 2M(t)} \cr
&& +{M(t)\over 8}\left[A_0^2(t)-A_1^2(t)\right]x^2-a(t)[xp+px],
 \end{eqnarray}
with
 \begin{eqnarray}
 U_L&=&\exp\left[i{M\over 2\hbar}\left({\dot{\rho}\over \rho}+2a\right)
   x^2\right]
   \exp\left[-{i\over 2\hbar}\ln\left( \sqrt{w_c \over \Omega}\rho\right)
     (xp+px)\right]\cr
 &&\times \exp\left[{i\over 2\hbar}(t -{\tau\over w_c})\left(p^2+w_c^2
    x^2\right)\right].
 \end{eqnarray}
For ${1 \over 4}(A_0^2-A_1^2)-4a^2+{2\over M}{d\over dt} (Ma)>0$,
the relation of Eq.~(17) becomes the relation between a general
quadratic system and a simple harmonic oscillator
\cite{SongUni,Seleznyova}. For ${1 \over
4}(A_0^2-A_1^2)-4a^2+{2\over M}{d\over dt} (Ma)\leq 0$, one may
find that the relation in Eq.~(17) is true, though, in these
cases, $U_L$ may not be useful in finding wave functions for a
general quadratic system which are localized for all time $t$.
Since $L_i$ and $K_i$ share the same algebraic structure, proving
Eq.~(17) constitutes a proof of the general relation of Eq.~(10).

It should be, however, mentioned that, even for the quadratic
systems of $C=-(3/16)I$, Eq.~(10) is more general than Eq.~(17),
as much as the realization of the algebra is not unique. For
example, generators $\tilde{L}_0,~\tilde{L}_1,~\tilde{L}_2$  of
the $SU(1,1)$ group can be realized as
 \be
 \tilde{L}_0=L_0,~~~\tilde{L}_1=-L_1,~~~\tilde{L}_2=-L_2.
 \ee
In this realization, Eq.~(10) is written as
 \begin{eqnarray}
 U_{\tilde{L}}\left( -i\hbar{\partial \over \partial t}
 + {1\over 2}(p^2+w_c^2x^2)\right)U_{\tilde{L}}^\dagger &=&
 -i\hbar{\partial \over \partial t}
 + {{M\over 8w_c^2}\left[A_0^2(t)-A_1^2(t)\right]p^2 }
   +{w_c^2 \over 2M(t)}x^2 \cr &&
   +a(t)[xp+px],
 \end{eqnarray}
with
 \begin{eqnarray}
 U_{\tilde{L}}&=&\exp\left[i{M\over 2\hbar w_c^2}\left({\dot{\rho}\over \rho}+2a\right)
   p^2\right]
   \exp\left[{i\over 2\hbar}\ln\left( \sqrt{w_c \over \Omega}\rho\right)
     (xp+px)\right]\cr
 &&\times \exp\left[{i\over 2\hbar}(t -{\tau\over w_c})\left(p^2+w_0^2
    x^2\right)\right].
 \end{eqnarray}

\section{Wave functions of the SU(1,1) systems}
Making use of the Baker-Campbell-Hausdorff (or, disentanglement)
formula [see, e.g., Ref.~\cite{Eberly}] with the commutation
relations in Eq.~(13), one can find that the operator $U$ is
written as
 \be
 U=\left(e^{\xi K_+}e^{\gamma K_0} e^{-\bar\xi K_-}\right)
    e^{i\varphi K_0},
 \ee
 where
 \begin{eqnarray}
 \xi&=&{-{\Omega \over \rho}+w_c\rho+iM(\dot{\rho}+2a\rho) \over
         {\Omega \over \rho}+w_c\rho- iM(\dot{\rho}+2a\rho)}
         e^{2iw_c t}, \\
 \gamma&=&\ln(1+|\xi|^2), \\
 \varphi&=&2(w_ct-\tau)
      -i \ln{ {\Omega \over \rho}+w_c\rho+iM(\dot{\rho}+2a\rho) \over
              {\Omega \over \rho}+w_c\rho-iM(\dot{\rho}+2a\rho) }.
 \end{eqnarray}
In Eq.~(22), $\bar\xi$ denotes the complex conjugate of $\xi$. An
element $g$ of the $SU(1,1)$ may be written in the form
\begin{equation}
g(\alpha,\beta)= \left(
\begin{array}{cc}
\alpha & \beta \\
\bar\beta & \bar\alpha
\end{array}\right),~~~~|\alpha|^2-|\beta|^2=1.
\end{equation}
Making use of  the realization of the generators
 \be
 K_0=\left(
    \begin{array}{cc}
    1/2 & 0  \\
    0 &-1/2
    \end{array}\right),~~~
 K_+=\left(
    \begin{array}{cc}
    0 & 1 \\
    0 & 0
    \end{array}\right),~~~
 K_-=\left(
    \begin{array}{cc}
    0 & 0  \\
    -1 & 0
    \end{array}\right),
 \ee
one can find that $g(\alpha,\beta)$ is parameterized as
 \begin{eqnarray}
\alpha &=&{e^{i\varphi/ 2}\over \sqrt{1-|\xi|^2}}=
               {e^{i(w_c t-\tau)}\over 2}
         \left[ {1\over \rho}\sqrt{\Omega \over w_c} +\rho\sqrt{w_c \over \Omega}
                +{iM\over
                \sqrt{w_c\Omega}}(\dot{\rho}+2a\rho)\right], \\
\beta &=&{\xi e^{-i{\varphi/ 2}}\over \sqrt{1-|\xi|^2}}=
          {e^{i(w_c t+\tau)}\over 2}
         \left[ -{1\over \rho}\sqrt{\Omega \over w_c} +\rho\sqrt{w_c \over \Omega}
                +{iM\over
                \sqrt{w_c\Omega}}(\dot{\rho}+2a\rho)\right].
 \end{eqnarray}

 Among the representations of SU(1,1) group, we only consider the
 UIRs \cite{Bargmann}. In a UIR, a basis
 state can be denoted as $|m,q_0,k>$ satisfying
 \begin{eqnarray}
 &&C\left(e^{-2i(m+q_0)t}|m,q_0,k>\right)=k(k-1)\left(e^{-2i(m+q_0)t}|m,q_0,k>\right), \\
 &&K_0\left(e^{-2i(m+q_0)t}|m,q_0,k>\right)=(m+q_0)\left(e^{-2i(m+q_0)t}|m,q_0,k>\right).
 \end{eqnarray}
There are four classes of UIRs, and $m$ must be integers
\cite{Bargmann,IKG}. When a group element is acted on a basis
state of a UIR, if we assume the completeness of the
representation, the result should be written as a linear
combination of the basis states of the UIR. Since
$e^{-2i(m+q_0)}|m,q_0,k>$ satisfies the Scr\"{o}dinger eqation
 \be
 i\hbar{\partial \over \partial t}
 \left( e^{-2i(m+q_0)t}|m,q_0,k>\right)=H_0 \left( e^{-2i(m+q_0)t}|m,q_0,k>\right),
 \ee
from the unitary relation of Eq.~(10), one can find that the state
given by
 \be
 |\Psi_{m,q_0,k}>=U\left(e^{-2i(m+q_0)t}|m,q_0,k>\right)
 \ee
should satisfy the Schr\"{o}dinger equation
 \be
i\hbar{\partial \over \partial t}
|\Psi_{m,q_0,k}>=H|\Psi_{m,q_0,k}>,
 \ee
 while $|\Psi_{m,q_0,k}>$ may be written as
 \be
 |\Psi_{m,q_0,k}>=V(g)\left(e^{-2i(m+q_0)t}|m,q_0,k>\right)=
 \sum_{m'}V_{m',m}^{(k,q_0)}(\alpha,\beta)\left(e^{-2i(m'+q_0)t}|m',q_0,k>\right).
 \ee

Though Eq.~(35) is valid in any UIR, from now on we only consider
the  representation of positive discrete series $D^+(k)$ where
$k>0,~q_0=k,~ k~{\rm is ~real},~{\rm and} ~m=0,1,2,3,\cdots$.
Since $q_0=k$ in this representation, $q_0$ will be omitted or
replaced by $k$. A basis state of $D^+(k)$ could then be written
as
 \be
 \left(e^{-2i(m+k)t}|m,k>\right)=\sqrt{\Gamma(2k)\over
 m!\Gamma(m+2k)}\left(K_+\right)^m\left(e^{-2ikt}|0,k>\right).
 \ee
The explicit expression of $V_{m',m}^{(k)}(\alpha,\beta)$, the
Bargmann functions, are known in this case, and $|\Psi_{m,k}>$ is
written as \cite{Bargmann,IKG}
 \be
 |\Psi_{m,k}>=
 \sum_{m'=0}^\infty
 V_{m',m}^{(k)}(\alpha,\beta)\left(e^{-2i(m'+k)t}|m',q_0,k>\right).
 \ee
As shown in appendix, $ V_{m',m}^{(k)}$ can be given as
 \begin{eqnarray}
 V_{m',m}^{(k)}(\alpha,\beta)  &=&
   {\Gamma(m+m'+2k) \over \sqrt{m!\Gamma(m+2k)(m')!\Gamma(m'+2k)}} \cr
 &&\times
   {\bar\alpha}^{-m-m'-2k}\beta^{m'}\left(-\bar\beta\right)^m
   F[-m,-m';-m-m'-2k+1; {\alpha\bar\alpha \over \beta\bar\beta}],
 \end{eqnarray}
where $F[a,b;c;z]$ is the hypergeometric function
\cite{RussianTable}.

\section{Generalized Harmonic Oscillators}
As is well-known \cite{Perelbook}, the representation spaces of
$k=1/4$ and $3/4$ of $D^+(k)$ reduce to the Hilbert space of a
simple harmonic oscillator. $|m,{1/ 4}>$ in a representation space
of the $SU(1,1)$ corresponds to $|2m>$ of a simple harmonic
oscillator which is an eigenstate of the Hamiltonian $H_0=2\hbar
w_cL_0$ with the energy eigenvalue $(2m+{1\over 2})\hbar w_c$. For
$k={3/ 4}$, $|n,3/4>$ corresponds to the eigenstate $|2m+1>$ of
$H_0$ with the energy eigenvalue $(2m+1+{1\over 2})\hbar w_c$.
Since the unitary relation of Eq.~(10) reduces to the one for the
quadratic systems if we choose a basis of the $su(1,1)$ algebra as
in Eq.~(16), for ${1 \over 4}(A_0^2-A_1^2)-4a^2+{2\over M}{d\over
dt} (Ma)>0$, one can find the explicit expressions of
$e^{-i(2m+{1\over 2})t}<x|U_L|m,{1\over 4}>$ and $e^{-i(2m+{3\over
2})t}<x|U_L|m,{3\over 4}>$, as in Ref.~\cite{SongUni,Song99}.

In this section, $V_{m',m}^{(k)}$ will be studied in more detail
for both cases of $k=1/4$ and $3/4$, which will generalize the
known results \cite{Seleznyova,Perelbook}. It will also be shown
that, if the unitary relation becomes a relation between the same
system described by $H_0$, the operator $U$ and (thus the
corresponding Bargmann function) can be written in a very simple
form.

\subsection{For a general quadratic system}
Making use of the transformation formula \cite{RussianTable}
 \be
  F[a,b;c;z]=(1-z)^{-a}F[a,c-b;c;{z\over z-1}],
  \ee
and Eq.~(A.2), one can find that $V_{m',m}^{(k)}(\alpha,\beta)$
may be written as
 \begin{eqnarray}
V_{m',m}^{(k)}(\alpha,\beta)
 &=&{\Gamma(m+m'+2k)\over \sqrt{m!\Gamma(m+2k)(m')!\Gamma(m'+2k)}}
    \left(\bar\alpha\right)^{-m-m'-2k}\left(\beta\right)^{m'-m}
    \cr
 & &\times  F[-m,-m-2k+1;-m-m'-2k+1; \alpha\bar\alpha]
 \cr
 &=&{(-)^m \over (2k-1)!}
    \sqrt{\Gamma(m+2k)\Gamma(m'+2k) \over m!(m')!}
    \left( \bar\alpha \right)^{-m'-2k}\alpha^m\beta^{m'-m} \cr
 & &\times F[-m,m+2k;2k;{1\over \alpha\bar\alpha}].
 \end{eqnarray}
For $k=1/4$, a relation between the hypergeometric function and
the associate Legendre function with non-negative integers $p,q$,
 \be
 F[-p,q+{1\over 2};{1\over 2};x^2]=(-)^q{(2p)!! \over (2q-1)!!}
   \left(1-x^2\right)^{p-q \over 2}P_{q+p}^{q-p}(x),
 \ee
obtained from a more general one in Ref.~\cite{RussianTable}, can
be used to find a simpler expression of
$V_{m',m}^{(1/4)}(\alpha,\beta)$, as
 \be
 V_{m',m}^{(1/4)}(\alpha,\beta)=
 {(-)^{m+m'}\over \sqrt{\bar\alpha}}\sqrt{(2m)!\over (2m')!}
 \left({\alpha\over \bar\alpha}\right)^{m+m' \over 2}
 \left({\bar\beta \over \beta}\right)^{m-m' \over 2}
 P_{m'+m}^{m'-m}\left({1 \over \sqrt{ \alpha\bar\alpha}} \right).
 \ee
For $k=3/4$, the formula
 \be
 F[-p,q+{3\over 2};{3\over 2};x^2]=(-)^q{(2p)!! \over (2q+1)!!}
 {1\over x}\left(1-x^2\right)^{p-q \over 2}P_{q+p+1}^{q-p}(x),
 \ee
can be used to find the fact that
 \be
 V_{m',m}^{(3/4)}(\alpha,\beta)=
 {(-)^{m+m'}\over \sqrt{\bar\alpha}}\sqrt{(2m+1)!\over (2m'+1)!}
 \left({\alpha\over \bar\alpha}\right)^{m+m'+1 \over 2}
 \left({\bar\beta \over \beta}\right)^{m-m' \over 2}
 P_{m'+m+1}^{m'-m}\left({1 \over \sqrt{ \alpha\bar\alpha}} \right).
 \ee

For the case of $M(t)=1$ and $a(t)=0$, with the choice of the
basis in Eq.~(16), $H$ of Eq.~(1) becomes the Hamiltonian of a
harmonic oscillator of unit mass and time-dependent frequency
($w(t)=\sqrt{{w_c \over 2}(A_0(t)+A_1(t))}$). In this case, one
can find that Eqs.~(42,44) exactly reproduce Eq.~(83) of the
Ref.~\cite{Seleznyova}.

\subsection{For a simple harmonic oscillator}
For the case of $A_0(t)=2w_c$ and $A_1(t)=a(t)=0$, $H$ of Eq.~(1)
becomes $H_0$ of Eq.~(3), and the unitary relation given in
Eq.~(10) becomes a relation between the same system. In this case,
$\rho$ satisfies the equation
 \be
 \ddot{\rho}-{\Omega^2 \over \rho^3}+w_c^2\rho=0,
 \ee
which makes it possible to analyze the unitary operator $U$ in
more detail. Making use of the fact in Eq.~(45), one can find that
 \be
 {d \over dt} \ln\xi=0,
 \ee
 and
 \be
 {d \over dt} \varphi=0.
 \ee

Though Eqs.~(46,47) are valid for general $k$, to be explicit, we
first consider the realization given in Eq.~(16). In this
realization, by defining
 \be
 a_c={1\over\sqrt{2\hbar w_c}}(w_cx+ip), ~~~
 a_c^\dagger={1\over\sqrt{2\hbar w_c}}(w_cx-ip),
 \ee
with a real constant $\varphi_c$ and a complex constant $\xi_c$,
one can find that $U_L$ can be written as
 \begin{eqnarray}
 U_L&=&\exp\left[{\xi_c a_c^\dagger a_c^\dagger \over 2}e^{-2iw_ct}\right]
 \exp\left[{\ln(1+|\xi_c|^2)\over 2}(a_c^\dagger a_c+ {1\over 2})\right]
 \exp\left[-{\bar\xi_c a_c a_c \over 2}e^{2iw_ct}\right]    \cr
&&\times
 \exp\left[{i\varphi_c \over 2}(a_c^\dagger a_c+{1\over 2} )\right].
 \end{eqnarray}
$U_L$ of $k=1/4$ or $3/4$, thus, shows that, if $a_c^\dagger
a_c^\dagger$~ ($a_ca_c$) is applied on a state to give a new
state, the phase factor $e^{-2iw_ct}$~ ($e^{2iw_ct}$) should be
multiplied at the same time, which proves that
 \be
 |\Psi_{m,k}>=
 \sum_{m'=0}^\infty c_{m,m'}\left(e^{-2i(m'+k)t}|m',k>\right),
 \ee
where $c_{m,m'}$ is a constant.

For a general $k$ with $A_0(t)=2w_c$, $A_1(t)=a(t)=0$, it may be
easy to find that wave functions satisfying Eq.~(34) should also
be written as in Eq.~(50). A wave function in a simple harmonic
oscillator system can be obtained by superposing the wave
functions of $k=1/4$ and $k=3/4$, so a wave function in this
system is written as
 \be
|\psi>=\sum_{n=0}^\infty c_n e^{-i(n+{1\over 2})w_c t}|n>,
 \ee
where $c_n$ is a constant.

\section{Harmonic oscillator with an inverse-square interaction}
It has been known that generators for $D^+(k)$ of the $SU(1,1)$
can be realized as \cite{inverse}
\begin{eqnarray}
 \hbar L_0^k &=&{1\over 4w_c}\left( p^2+ {2g\over x^2}\right)+{w_cx^2 \over 4}, \cr
 \hbar L_1^k &=&-{1\over 4w_c}\left( p^2+ {2g\over x^2}\right)+{w_cx^2 \over 4},\cr
 \hbar L_2^k &=& -{1\over 4}(xp+px),
\end{eqnarray}
where $g=2(k-{1\over 4})(k-{3\over 4})$. For the system of a
Hamiltonian $H_k=2w_c L_0^k$ on the half-line $x>0$, the wave
functions are given as \cite{Calogero}
 \begin{eqnarray}
 \phi_n^{s-} (k;x,t) &=&\left({4w_c\over \hbar}\right)^{1/4}
  \left( {n!\over \Gamma(n+2k)}\right)^{1/2} \cr
 &&\times e^{-2i(n+k)w_c t}
  \left(w_c x^2\over \hbar\right)^{k-{1\over 4}}
  \exp\left(-{w_cx^2 \over 2\hbar}\right)
  L_n^{2k-1}\left({w_c x^2 \over \hbar}\right),
 \end{eqnarray}
 where $L_n^\alpha$ is the associated Laguerre polynomial defined
through the equation
 \be
 x{d^2 L_n^\alpha \over dx^2}+(\alpha+1-x){dL_n^\alpha \over dx}
 +nL_n^\alpha(x)=0.
 \ee
However, the fact
 \[\phi_n^{s-} ({1\over 4};x,t)=(-)^n
 \left({2 \sqrt{w_c}\over 2^{2n}(2n)!\sqrt{\pi\hbar}}\right)^{1/2}
 e^{-2i(n+{1\over 4})w_c t}\exp\left(-{w_cx^2 \over 2\hbar}\right)
 H_n \left(\sqrt{w_c\over \hbar}x\right),\]
implies that
 \be
 \phi_n^s (k;x,t)\equiv e^{-2i(n+k)w_c t}<x|n,k>=
 (-)^n \phi_n^{s-}(k;x,t).
 \ee

If we choose $L_0^k,~L_1^k,~L_2^k$ as the generators of the
$SU(1,1)$, $U$ becomes $U_L$, and $U_L\phi_n^{s-} (k;x,t)$ can be
calculated as \cite{SongUni}
 \begin{eqnarray}
\phi_n^- (k;x,t)&=& U_L\phi_n^{s-} (k;x,t) \cr
 &=&\left({4\Omega \over \hbar\rho^2}\right)^{1/4}
  \left( {n!\over \Gamma(n+2k)}\right)^{1/2} e^{-2i(n+k)\tau}
  \left(\Omega x^2\over \hbar\rho^2\right)^{k-{1\over 4}}   \cr
 & &\times \exp\left[-{x^2 \over 2\hbar}
     \left({\Omega\over \rho^2}-iM{\dot{\rho}\over \rho}-2iMa\right)\right]
  L_n^{2k-1}\left({\Omega x^2 \over \hbar\rho^2}\right).
  \end{eqnarray}
Eqs.~(37) and (55) then suggest that
 \be
 \phi_n^-
 (k;x,t)=\sum_{m=0}^\infty(-)^{n+m}V_{m,n}^{k}(\alpha,\beta)\phi_m^{s-}(k;x,t).
 \ee
Making use of the integration formula \cite{EMOT},
 \begin{eqnarray}
 &&\int_0^\infty e^{-bx}x^\alpha L_n^\alpha(\lambda x)L_m^\alpha
 (\mu x) dx \cr
 &&={\Gamma(m+n+\alpha+1) \over m!n!}
    {(b-\lambda)^n (b-\mu)^m \over b^{m+n+\alpha+1}}
    F[-m,-n;-m-n-\alpha;
      {b(b-\lambda-\mu)\over (b-\lambda)(b-\mu)}]
 \end{eqnarray}
which is valid for ${\rm Re}~ \alpha >-1$ and ${\rm Re}~ b>0$, one
can indeed find that
 \be
 \int_0^\infty \bar{\phi}_m^{s-}(k;x,t)\phi_n^- (k;x,t) dx
 =(-)^{n+m}V_{m,n}^{k}(\alpha,\beta).
 \ee

If we consider a system described by the Hamiltonian
$H_k(\epsilon)= {1\over 2(1+i\epsilon)}\left( p^2+ {2g\over
x^2}\right)+(1+i\epsilon){w_c^2x^2 \over 2}$ with real positive
$\epsilon$, one can show that the kernel (propagator) $K(x_b,
t_b;x_a, t_a)$ of the system (see, e.g., Ref.~\cite{IKG}) reduces
to the kernel of free particle of unit mass in the limit of
$t_b\rightarrow t_a+0$ and $\epsilon\rightarrow 0$, which would
imply completeness of the set $\{\phi_n^{s-} (k;x,t)|
~n=0,1,2,\cdots\}$. Indeed, if we assume completeness, the fact in
Eq.~(59) amounts to a proof for the relation in Eq.~(57).

One can also take $L_0^k,~\tilde{L}_1^k,~\tilde{L}_2$ as the
generators of the $SU(1,1)$, while
 \be
\tilde{L}_1^k=-L_1^k,~~~\tilde{L}_2^k=-L_2^k.
 \ee
Since
$\tilde{L}_+^k=e^{-2iw_ct}(\tilde{L}_1^k+i\tilde{L}_2^k)=-L_+^k$,
Eqs.~(36,55) imply that
 \be
 \phi_n^{s-} (k;x,t)=\sqrt{\Gamma(2k)\over m!\Gamma(m+2k)}
 \left(\tilde{L}_+^k\right)^m \phi_0^{s-} (k;x,t).
 \ee
If we use these generators in the unitary relation of Eq.~(10),
the relation and Eq.~(37) imply that the wave function
 \be
 \phi_n
 (k;x,t)=\sum_{m=0}^\infty V_{m,n}^{k}(\alpha,\beta)\phi_m^{s-}(k;x,t),
 \ee
satisfies the Schr\"{o}dinger equation
 \begin{eqnarray}
 &&i\hbar{\partial \over \partial t}\phi_n(k;x,t) \cr
 &&=\left[{M\over 8w_c^2}\left(A_0^2(t)-A_1^2(t)\right)
 \left(-\hbar^2{\partial^2 \over \partial x^2}
   +{2g\over x^2}\right)  \right]\phi_n(k;x,t)  \cr
 &&~~~   +\left[{w_c^2 \over 2M(t)}x^2
   -ia(t)\hbar(2x{\partial \over \partial x}+1)\right]\phi_n
 (k;x,t).
 \end{eqnarray}

\section{Discussions}
We have shown that, for the systems of $su(1,1)$ symmetry, there
is a unitary relation between the system whose Hamiltonian is
given as a linear combination of the generators of $SU(1,1)$ group
with time-dependent coefficients and a system of the Hamiltonian
which is simply proportional to the generator of the compact
subgroup. The unitary relation is obtained through an extension of
that between the general quadratic system and a simple harmonic
oscillator. However, it should be mentioned that the relation is
still formal, in the sense that the explicit form of the relation
is given providing the classical solutions $u(t),v(t)$ are known.
For the case that $M$ is constant and $a=0$, if $A_0^2<A_1^2$, Eq.
(5) becomes the equation of motion of an inverted harmonic
oscillator, so that $\rho$ diverges as time goes to infinity. If
$\rho$ diverges, for a quadratic system, the probability
distribution of a wave function obtained through the unitary
relation spreads out all over the space, while it may be possible
that a meaningful system could be defined algebraically with
diverging $\rho$.

Another point worthy of being mentioned is that the formal
relation is true even for the case of negative $M(t)$. In fact,
for a constant Hamiltonian $H$, the Schr\"{o}dinger equation is
invariant under the exchange of $H\leftrightarrow -H$ and
$t\leftrightarrow -t$. In the case of $A_0(t)=-2w_c$ and
$A_1(t)=a(t)=0$ where $M(t)=-1$ and thus $H=-H_0$, one can take
the classical solutions as $u(t)=\sin w_ct,~v(t)=\cos w_ct$, so
that $U=-\exp[2it(2w_cK_0)]$. By applying this $U$ on a stationary
state $e^{-2i(m+q_0)w_ct}|m,q_0,k>$, one will have the sate
$-e^{2i(m+q_0)w_ct}|m,q_0,k>$. This fact, therefore, suggests that
the invariance may be included in the unitary relation.

It would be interesting to find similar unitary relations in the
systems with other symmetries. The unitary relation for the
$SU(1,1)$ system has been found based on the relation in harmonic
oscillators which may be the simplest system with the symmetry.
This imply that, if we find a unitary relation in a simple system
of a symmetry, the relation could be generalized for other systems
with the same symmetry. Though the $SU(1,1)$ is a non-compact
group, the generalization itself would be possible for a compact
group. In addition, it would be interesting to find the
implications of the unitary relation in a system where the
$su(1,1)$ symmetry is a part of the symmetry of the system.

\begin{acknowledgments}
The author thanks John Klauder and Physics Department of
University of Florida for the hospitality and discussions. This
work was supported in part by the Korea Research Foundation Grant
(KRF-2002-013-D00025) and by NSF Grant 1614503-12.
\end{acknowledgments}

\appendix*
\section*{Appendix}
\setcounter{equation}{0}

In order to show that the Bargmann function given in Eq.~(38) is
equal to the standard expression \cite{Bargmann,IKG}, the
hypergeometric function in the equation can be written as
 \begin{eqnarray}
 &&F[-m,-m';-m-m'-2k+1;{\alpha\bar\alpha \over \beta\bar\beta}]
    =F[-m,-m';-m-m'-2k+1;1+{1 \over \beta\bar\beta}]\cr
 &&={\Gamma(m'+2k)\Gamma(m+2k) \over\Gamma(2k)\Gamma(m+m'+2k)}
 F[-m,-m';2k;-{1\over \beta\bar\beta}],
 \end{eqnarray}
where the last equality is obtained through a formula given in
Ref.~\cite{RussianTable}. With the Appell's symbol $(a,s)$ defined
for non-negative integers $s$ by
 \[ (a,s)\equiv \left\{\begin{array}{cc}
     1  & ~~~(s=0) \\
  a(a+1)\cdots(a+s-1) &~~~(s>0)
   \end{array}\right\},
 \]
a formula
 \be
 F[-l,b;c;-y]={(b,l) \over (c,l)}y^l F[-l,1-l-c;1-l-b; -1/y],
 \ee
is known for a non-negative integer $l$ in Ref.~\cite{Bargmann},
which is valid as long as $(b,l)\neq 0$. For $m\geq m'$, Eq.~(A.2)
can be used to find
 \begin{eqnarray}
 &&F[-m,-m';-m-m'-2k+1;{\alpha\bar\alpha \over \beta\bar\beta}]\cr
 &&={m!\Gamma(m+2k) \over (m-m')!\Gamma(m+m'+2k)}
    (-\beta\bar\beta)^{-m'}
    F[-m',-m'-2k+1;1-m'+m;-\beta\bar\beta].~~
 \end{eqnarray}
For $m'\geq m$, Eq.~(A.2) can also be used to give
 \begin{eqnarray}
 &&F[-m,-m';-m-m'-2k+1;{\alpha\bar\alpha \over \beta\bar\beta}]
 =F[-m',-m;-m-m'-2k+1;{\alpha\bar\alpha \over \beta\bar\beta}]\cr
 &&={(m')!\Gamma(m'+2k) \over (m'-m)!\Gamma(m+m'+2k)}
    (-\beta\bar{\beta})^{-m}
    F[-m,-m-2k+1;1-m+m';-\beta\bar\beta].~~~
 \end{eqnarray}

After some algebra with the above formulas, one can find that, the
Bargmann function of $D^+(k)$ given in Eq.~(38) is equivalent to
the standard expression \cite{Bargmann,IKG}:\newline for $m'\geq
m$,
 \be
 V_{m',m}^{(k)}(\alpha,\beta)=A_{m',m}\left(\bar{\alpha}\right)^{-m'-m-2k}
 \left(\beta\right)^{m'-m} F(-m,1-m-2k;1+m'-m;-\beta\bar\beta)
 \ee
and for $m'\leq m$,
 \be
 V_{m',m}^{(k)}(\alpha,\beta)=A_{m,m'}\left(\bar{\alpha}\right)^{-m'-m-2k}
    \left(-\bar\beta\right)^{m-m'}
 F(-m',1-m'-2k;1+m-m';-\beta\bar\beta),
 \ee
where
 \be
 A_{m'm}={1\over (m'-m)!}\left({(m')!\Gamma(m'+2k) \over
 m!\Gamma(m+2k)}\right)^{1\over 2}.
 \ee

Since the hypergeometric series of any hypergeometric function
used in this paper terminates, the series always converges. Making
use of the expression of $U$ given in Eq.~(22) and the basis state
in Eq.~(36), the expression of Bargmann function (Eqs.~(A.5-6))
can also be directly derived for $D^+(k)$ as
\begin{eqnarray}
 V_{m',m}^{(k)}(\alpha,\beta)&=&e^{-2i(m-m')t}<m',k|U|m,k> \cr
 &=&e^{i[(m+k)\varphi-2(m-m')t]}
   <m',k|e^{\xi K_+}e^{\gamma K_0}e^{-\bar\xi K_-}|m,k> \cr
 &=&e^{i[(m+k)\varphi-2(m-m')t]}
   <m',k|e^{-\bar\xi K_-}e^{-\gamma K_0}e^{\xi K_+}|m,k> \cr
 &=&{e^{i(m+k)\varphi}\Gamma(2k) \over
       \sqrt{(m')!m!\Gamma(m+2k)\Gamma(m'+2k)}}  \cr
 & &\times     \sum_{p,q=0}^\infty
       {(-\bar\xi)^p\xi^q \over p!q!}<0,k|(K_-)^{m'+p}e^{-\gamma K_0}
       (K_+)^{q+m}|0,k>,
\end{eqnarray}
while the remained procedures are straightforward.



\begin{thebibliography}{}
\bibitem{comment} In fact, in proving the relation, it is more convenient to use
the commutation relations given in Eq.~(2).
\end{thebibliography}
{}
\end{document}